\documentclass[aps,prl,twocolumn,amsmath,showpacs,groupedaddress]{revtex4}  
\usepackage{graphicx}  
\usepackage{dcolumn} 
\usepackage{bm}    
\usepackage{amssymb}

\begin{document}

\title{Rounding by disorder  of  first-order quantum phase transitions: emergence of quantum critical points}
\author{Pallab Goswami}
\affiliation{Department of Physics and Astronomy, University of California Los Angeles, Los Angeles, California, 90095-1547}
\author{David Schwab}
\affiliation{Department of Physics and Astronomy,  University of
California Los Angeles, Los Angeles, California, 90095-1547}
\author{Sudip Chakravarty}
\affiliation{Department of Physics and Astronomy, University of California Los Angeles, Los Angeles, California, 90095-1547}
\date{\today}

\begin{abstract}
We give a heuristic argument for  disorder rounding of a first order quantum phase transition into a continuous phase transition. From both weak and strong disorder analysis of the the $N$-color quantum Ashkin-Teller model in one spatial dimension, we find that  for $N \geq 3$, the first order transition is rounded to a continuous transition and the physical picture is the same as the random transverse field Ising model for a limited parameter regime. The results are strikingly different from the corresponding classical problem in two dimensions where the fate of the renormalization group flows  is a fixed point corresponding to $N$-decoupled pure Ising models. 
\end{abstract}

\pacs{}
\maketitle 
The effect of disorder on continuous classical phase transitions has been intensely studied over many decades,~\cite{Harris:1974} but less is known about its effect on  first order  transitions. Imry and Wortis~\cite{Imry:1979} argued that arbitrarily weak disorder can actually round a classical first order transition into a continuous transition. Subsequently, Hui and Berker~\cite{Hui:1989} and Aizenmann and Wehr~\cite{Aizenman:1989,Aizenman:1990} have made important contributions to this topic. In light of numerous recent experiments involving quantum phase transitions in systems that  inevitably contain many sources of disorder, the issue of disorder rounding in first order quantum phase transition ($\mathrm{QPT_{1}}$) has acquired considerable urgency.   

There are two important questions: (a) Can disorder convert a $\mathrm{QPT_{1}}$  to a continuous one? (b) Might the theorem for a disordered classical system require modification in a quantum context? There is, as we shall see, a simple intuitive affirmative answer to (a), but (b) is more subtle because statics and dynamics are entangled in a  quantum phase transition.   To answer (b) and to explore more fully the issues involved, we consider below a model and study it in considerable detail. It is important to note that for even the classical random bond Potts model for $q>4$ and $d=2$, for which the pure system has a first order transition,  the critical exponents of the disordered system do not belong to the simple  universality class of the pure Ising model--- although the critical exponent $\nu\approx 1$, all other critical behavior is different~\cite{Olson:1999}. 

We will answer question (a) by  a heuristic argument patterned along  an argument by Berker~\cite{Berker:1993}, although we differ in our analysis for the case of continuous symmetry.  Let the disorder couple to the Hamiltonian in such a way that its symmetry is unchanged. For example, disorder may couple to a nearest  neighbor bond (generally to energy-like variables) without affecting the symmetry. 
In contrast,  a site random field  breaks the symmetry explicitly.  Consider changing  a generic  tuning parameter $g$ that may be the ratio of the amplitudes of two noncommuting terms in the Hamiltonian, which controls the quantum fluctuations and results in a first order transition. 

We would like to show that in  the presence of disorder coexistence of phases is not possible at the transition, and the quantum fluctuations do not have a scale. If this is true, and if the state corresponding to $g=0$ is still a broken symmetry state (this is why we imposed the specific requirement on disorder earlier) and the $g=\infty$ is the quantum disordered state, the conclusion must be that the transition has been converted to a quantum critical point.

The proof is by contradiction.  Assume that the first order transition is at $g_{c}$, implying coexistence of phases. However, in the presence of disorder there will be local fluctuations of  $g_{c}$.   Thus, within a putative quantum disordered region, randomness can nucleate an ordered region of linear dimension $L$, with a gain in the volume energy $\propto L^{d/2}$ (assuming central limit theorem), while the price in the surface energy is $\propto L^{d-1}$. The same is true for a putative ordered region. Therefore, for $d  <  2$ (discrete) , the picture is that of a ``domain within domain'', and there is no scale, as required for a $\mathrm{QPT_{1}}$.  In contrast, for $g < g_{c}$ nucleation of one broken symmetry phase within another does not gain any energy (disorder does not break the relevant symmetry), but the surface energy is increased. Therefore, by contradiction, coexistence of phases is not possible, and the transition at $g_{c}$ must  be continuous. We have tacitly assumed that the transition involves a broken symmetry. If this is not the case, there is no particular reason for a sharp transition to remain at $g_{c}$, and the disorder will simply smear out the transition.

The case of continuous symmetry is a little subtle. While one may like to argue that the domain wall energy is $L^{d-2}$~\cite{Berker:1993}, as in the Imry-Ma argument, this is typically incorrect. If the domain wall connects a broken symmetry state with an unbroken symmetry state, where the amplitude of the order parameter vanishes, the domain wall energy is still $L^{d-1}$, as in the case of discrete symmetry. However, if the $\mathrm{QPT_{1}}$ is effected by tuning a ``magnetic field'' that changes the state from one broken symmetry direction to another, the domain wall energy is indeed $L^{d-2}$, and the borderline dimensionality is $d=4$. Of course from Mermin-Wagner theorem there is no long range order in $d=2$ at any finite temperature, by either first order or continuous transitions; so, for the classical case the question is moot at $d=2$.

There are no simple arguments known to us for the borderline dimensionalities,  but from the rigorous version of the Imry-Ma argument for the random field case~\cite{Imbrie:1984,Aizenman:1989,Aizenman:1990}, it is safe to conjecture that the above argument should also hold for these cases, because there is a close connection between the two problems as noted by Imry~\cite{Imry:1984}. Note that the dynamic critical exponent $z$ does not enter in this analysis --- all we need is the extensivity of the ground state energy and its normal fluctuations in the thermodynamic limit. The principal disordering agent that washes out the coexistence is the fluctuations due to impurities and not quantum fluctuations. Quantum fluctuations can only help the process of smoothing the coexistence. Of course the fate of the system in dimensions higher than 2  must depend of the quantum fluctuations. In addition, the actual dynamics of the system must involve these fluctuations as well.

It is useful to note that, in contrast, the Harris criterion~\cite{Harris:1974} that determines the influence of impurities at a critical point, {\em inter alia} a quantum critical point,   does depend on the quantum dynamics. In order to see this, let us rephrase the Harris criterion along an argument essentially due to Mott~\cite{Mott:1981}. On one hand, disorder in a domain of linear dimension, $\xi$, the correlation length of the pure system defined by the decay rate of the equal time correlation function,  will give rise to fluctuations of the quantum critical point $g_{c}$ of fractional width $\Delta g \sim \xi^{-d/2}$. On the other hand, $\Delta g$ must be less than the reduced distance from the quantum criticality implied by $\xi$, that is $\sim \xi^{-1/\nu}$, for the criticality to remain unchanged, where $\nu$ is the correlation length exponent of the $d$-dimensional {\em quantum} system at zero temperature. Hence, $\nu>2/d$. Otherwise, the system may be described by a new disorder fixed point for which the same relation will apply with the replacement of the critical exponent of the pure system by the critical exponent of the new fixed point, as in the theorem of Chayes {\em et al}~\cite{Chayes:1986}. In either case the quantum dynamics is important because the relevant length scale close to the critical point is the {\em diverging correlation length}, $\xi$.  By contrast, the argument involving $\mathrm{QPT_{1}}$ is restricted by a {\em finite correlation length}, hence the balance is between the volume energy and the surface energy of a fluctuating domain nucleated by the impurities.

We now turn to the question (b) and study, using both a perturbative renormalization group and a real space decimation procedure~\cite{Ma:1979,Dasgupta:1980,Fisher:1995}, a one-dimensional quantum spin chain, the random $N$-color Ashkin-Teller model~\cite{Grest:1981,Fradkin:1984,Shankar:1985} in the presence of disorder.  We consider the regime in which the pure model has a $\mathrm{QPT_{1}}$. The corresponding classical problem in two spatial dimensions where the quenched disorder is isotropic, the renormalization group flows curl back to the pure decoupled Ising fixed point~\cite{Cardy:1996,Pujol:1996}, at least for weak couplings.

The Hamiltonian of the system is~\cite{Kohmoto:1981}
\begin{eqnarray}
H=-\sum_{\alpha=1}^N \sum_{i=1}^{L} \left(J_i\, \sigma^{(\alpha)}_{3,i} \sigma^{(\alpha)} _{3,i+1}
+h_i \, \sigma^{(\alpha)}_{1,i}\right)
\nonumber \\
-\epsilon \sum_{\alpha < \beta}^N \sum_{i=1}^{L}  \left(J_i\, \sigma^{(\alpha)}_{3,i} \sigma^{(\alpha)} _{3,i+1}
 \sigma^{(\beta)}_{3,i} \sigma^{(\beta)} _{3,i+1} +h_i \, \sigma^{(\alpha)}_{1,i}\sigma^{(\beta)}_{1,i}\right)
 \label{eq:Hamiltonian}
\end{eqnarray}
Here, Latin letters index lattice sites and Greek letters label colors and the $\sigma$'s are the Pauli opertaors.  The $J_i$ and $h_i$ are random variables taken from a distribution restricted to only positive values, while $\epsilon$ is a disorder independent positive constant.  For the random transverse field Ising model (RTFIM, $\epsilon=0$), a local gauge transformation may be performed to make all couplings positive, so the original couplings can take negative values.  The coupling between colors destroys this freedom so we must restrict ourselves to positive couplings from the outset.  We have parametrized the system  so that the intercolor couplings are proportional to the bond or field at that site.  We also restrict ourselves to $\epsilon \geq 0$.  Note the invariance of  the Hamiltonian with respect to the following duality transformations:
$\sigma^{(\alpha)}_{3,i} \sigma^{(\alpha)} _{3,i+1} \to \mu_{1,i}^{\alpha}$,
$\sigma^{(\alpha)}_{1,i} \to \mu^{(\alpha)}_{3,i} \mu^{(\alpha)} _{3,i+1}$,
$J_{i} \to h_{i}$,
where $\mu$'s are the dual Pauli operators. For the uniform system and with $N \geq 3$ and for $\epsilon>0$, there is a first order transition from a paramagnetic to an ordered state \cite{Grest:1981,Fradkin:1984,Shankar:1985,Ceccatto:1991}.

For weak disorder and weak four spin coupling, we can consider the continuum action in terms of Majorana fermions. The random 
$N$-color quantum Ashkin-Teller model can be described by an $O(N)$ Gross-Neveu model (GNM) with random mass.~\cite{Dotsenko:1985} The GNM action is given by
\begin{eqnarray}
S&=&\frac{v_{F}}{2}\int_{x,\tau} \bigg[ \sum_{\alpha=1}^{N}\bar{\psi}^{(\alpha)}\bigg(\frac{1}{v_{F}}\partial_{\tau}\sigma_{3}+\partial_{x}\sigma_{1}+m(x,\tau)\bigg)\psi^{(\alpha)} \nonumber \\
& &-\frac{g}{2}\left(\sum_{\alpha=1}^{N}\bar{\psi}^{(\alpha)}\psi^{(\alpha)}\right)^{2}\bigg],
\end{eqnarray}
where $v_{F}$ is the Fermi velocity, $\int_{x,\tau} \equiv \int dx d\tau$, and $\psi^{(\alpha)}=[\psi^{(\alpha)}]^{\dagger}$, $\bar{\psi}^{(\alpha)}=[\psi^{(\alpha)}]^{T}i\sigma_{2}$. In the above equation the random mass $m(x,\tau)$ follows Gaussian white noise distribution such that the correlation is given by 
$\overline{m(x,\tau)m(x^{'},\tau^{'})}=\Delta \delta(x-x^{'}).$
After averaging over disorder using $n$-replicas, we obtain
\begin{eqnarray}
\overline{S}&=&\frac{v_{F}}{2}\int_{x,\tau}\sum_{\alpha=1}^{N}\sum_{a=1}^{n}\bar{\psi}_{a}^{(\alpha)}\bigg(\frac{1}{v_{F}}\partial_{\tau}\sigma_{3}+\partial_{x}\sigma_{1}\bigg)\psi_{a}^{(\alpha)} \nonumber \\
& &-\frac{gv_{F}}{2}\int_{x,\tau} \sum_{a=1}^{n}\left(\sum_{\alpha=1}^{N}\bar{\psi}_{a}^{(\alpha)}\psi_{a}^{(\alpha)}\right)^{2}  
\nonumber \\ 
& &-\frac{\Delta v_{F}^{2}}{2}\int_{x,\tau}\int_{x',\tau'}\sum_{\alpha,\gamma}^{N}\sum_{a,b}^{n}\bar{\psi}_{a}^{(\alpha)}(x,\tau)\psi_{a}^{(\alpha)}(x,\tau)\nonumber \\
& &\delta(x-x')\bar{\psi}_{b}^{(\gamma)}(x',\tau^{'})\psi_{b}^{(\gamma)}(x',\tau^{'}).
\end{eqnarray}
Here the index $a$ corresponds to replicas. Simple power counting shows that $\Delta$ is a relevant operator with scaling dimension $1$ and $g$ is a marginal operator.  For even number of colors $(N=2M)$, this action can be expressed in terms of $M$ Dirac fermions and can be bosonized~\cite{Giamarchi:2004}; for $N=2M+1$ there will be a leftover Majorana fermion, which makes the analysis more complex but physical answers are the same.  A perturbative renormalization group calculation following Giamarchi and Schulz\cite{Giamarchi:1988} up to $O(\tilde{\Delta})$, $O(\tilde{\Delta} g)$ and $O(g^{2})$ lead to the following recursion relations, where $\tilde{\Delta}=\Delta a$, $a$ being the lattice spacing:
\begin{eqnarray}
\frac{d \tilde{\Delta}}{d l}&=&\tilde{\Delta} +\frac{(2M-1)}{\pi}\tilde{ \Delta} g, \\
\frac{d g}{d l}&=&\frac{(M-1)}{\pi}g^{2}+\frac{\tilde{\Delta} g}{3\pi} ,\\
\frac{d v_{F}}{d l}&=&\bigg[z-1-\frac{\tilde{\Delta}}{3\pi}(1-\frac{g}{\pi})\bigg]v_{F},
\end{eqnarray}
where $z$ is the dynamic exponent. These recursion relations are valid for $g/\pi \ll1$. For $v_{F}$ to remain fixed, $z$ must vary continuously:
\begin{equation}
z=1+\frac{\tilde{\Delta}}{3\pi}(1-\frac{g}{\pi}).
\end{equation}
There is one unstable fixed point $\Delta^{*}=g^{*}=0$ (free fermion fixed point with $z=1$) and both disorder and four spin coupling constants are relevant perturbations at this fixed point and flow away to the strong coupling regime. This makes it necessary to attack the problem using strong disorder renormalization group approach. 

The above flow equations should be contrasted with the flow equations of the classical $N$-color Ashkin-Teller model with quenched disorder. In the classical case the the disorder averaged action is local and disorder as well as $g$ are marginal perturbations. For this reason one has to calculate up to $O(\Delta^{2})$, $O(g^2)$ and $O(\Delta g)$~\cite{Cardy:1996,Murthy:1987,Pujol:1996}. In the replica limit, $n\rightarrow0$, the flows curl around and end up at the decoupled Ising fixed point, at least at weak couplings. 

To solve the random system, we will now employ the strong disorder renormalization group technique~\cite{Ma:1979,Dasgupta:1980,Fisher:1995,Igloi:2005}.  There are some similarities with the analysis of the quantum $q$-state Potts chain for $q>4$ for which it was argued that this transition is described by the infinite disorder fixed point, at least when the disorder is strong~\cite{Senthil:1996}. 
The decimation equations can be inferred from the structure of the energy levels.  We will demonstrate the calculation for a site decimation and note that the bond decimation equations follow from duality.  If the magnetic field on site $i$ is the largest coupling, the unperturbed Hamiltonian is
\begin{eqnarray}
H_0=-h_i \sum_{\alpha=1}^{N} \sigma_{1,i}^{(\alpha)}-\epsilon h_i \sum_{\alpha<\beta} \sigma_{1,i}^{(\alpha)}\sigma_{1,i}^{(\beta)}
\end{eqnarray}
and the ground state is $\left|\rightarrow \rightarrow \cdots \rightarrow \right>$ with energy $E_0=-N h_i - {N \choose{2}} \epsilon h_i$.  The Hilbert space is written as a tensor product of the spins at site $i$ for each color.  There are $N$ first excited state of the form $\left|\rightarrow \leftarrow \rightarrow \cdots \right>$, where one color has its spin flipped.  Each of these states has energy $E_1=E_0+2 h_i + 2 (N-1)h_i \epsilon$.  In general, the $r^{th}$ energy level has ${N \choose{r}}$ states with $r$ colors flipped and energy $E_r=E_0+2 r h_i+2 r (N-r) h_i \epsilon$.  The coupling of site $i$ to the rest of the system is the perturbation part
\begin{eqnarray}
V=-J_{i-1} \sum_{\alpha=1}^{N} \sigma_{3,i-1}^{(\alpha)} \sigma_{3,i}^{(\alpha)}-J_{i} \sum_{\alpha=1}^{N} \sigma_{3,i}^{(\alpha)} \sigma_{3,i+1}^{(\alpha)}
\\
-\epsilon J_{i-1} \sum_{\alpha<\beta} \sigma^{(\alpha)}_{3,i-1} \sigma^{(\alpha)} _{3,i}
 \sigma^{(\beta)}_{3,i-1} \sigma^{(\beta)}_{3,i}\nonumber
 \\
 -\epsilon J_{i} \sum_{\alpha<\beta} \sigma^{(\alpha)}_{3,i} \sigma^{(\alpha)} _{3,i+1}
 \sigma^{(\beta)}_{3,i} \sigma^{(\beta)}_{3,i+1}\nonumber
\end{eqnarray}
Within second order degenerate perturbation theory, the Ising terms connect the ground state to the first excited states only, while the 4-spin couplings connect to the second excited states.  Thus, the $4^N$-fold degeneracy of the ground state (due to the neighboring spins) is split by $V$:
\begin{eqnarray}
E_0'\simeq E_0 -\sum_{\alpha} \frac{\left(J_{i-1}\sigma_{3,i-1}^{(\alpha)} +J_{i}\sigma_{3,i+1}^{(\alpha)}\right)^2 }{2 h_i+2 h_i \epsilon (N-1)}\nonumber
\\
-\sum_{\alpha<\beta}\frac{\left( \epsilon J_{i-1} \sigma^{(\alpha)}_{3,i-1} \sigma^{(\beta)}_{3,i-1} +\epsilon J_i \sigma^{(\alpha)}_{3,i+1} \sigma^{(\beta)}_{3,i+1}\right)^2}{4 h_i+4 \epsilon h_i (N-2)}
\end{eqnarray}
The cross terms yield an effective Ising coupling between sites $i-1$ and $i+1$ given by
\begin{eqnarray}
\tilde{J}=\frac{J_{i-1}J_i}{h_i(1+\epsilon(N-1))}
\end{eqnarray}
and an effective four spin coupling
\begin{eqnarray}
\tilde{\epsilon}\tilde{J}= \frac{\epsilon^2}{2}\frac{J_{i-1}J_i}{h_i(1+\epsilon(N-2))}
\end{eqnarray}
Since $h_i$ was the largest coupling in the system, the new effective two- and four-spin couplings are both smaller than their original respective counterparts.  We see that the decimation results in new effective bonds and fields given by
\begin{equation}
\tilde{J_i}=\frac{J_iJ_{i+1}}{h_i\kappa}, \; 
\tilde{h}_i=\frac{h_ih_{i+1}}{J_i\kappa}, 
\label{eq:dual}
\end{equation}
where we have introduced
$\kappa=1+(N-1)\,\epsilon$,
and $\epsilon$ also renormalizes as
\begin{equation}
\tilde{\epsilon}=\frac{\epsilon^2 (1+(N-1)\epsilon)}{2\,(1+(N-2)\epsilon)}.
\end{equation}
Equations~(\ref{eq:dual})  exhibit the duality present in Eq.~(\ref{eq:Hamiltonian}) upon interchange of the couplings $h \leftrightarrow J$.  As long as $\epsilon$ is initially less than some $\epsilon_c(N)$, it is clear that $\epsilon$ will be reduced by this decimation.  

Let  $\Omega$ to be the largest energy scale still active in the problem and define  $\zeta=\ln{\Omega/J}$, $\beta=\ln{\Omega/h}$, and $\Gamma=\ln{\Omega_I/\Omega}$ where $\Omega_I$ is the strength of the original strongest bond.  
Then  the distributions of logarithmic bonds and fields at energy $\Gamma$, denoted by $P(\zeta)$ and $R(\beta)$, respectively, satisfy a set of flow equations. (We suppress the dependence of $P$ and $R$ on $\Gamma$.)  For $P(\zeta)$ we get

\begin{eqnarray}
\frac{\partial P}{\partial \Gamma} = \frac{\partial P}{\partial \zeta} &+& R(0) \int d \zeta ' P(\zeta ')
P(\zeta -\zeta '- \ln{\kappa})
\nonumber\\
 &+& \left(P(0)-R(0)\right)P (\zeta)
\end{eqnarray}
with a similar equation for flow of $R$ upon the replacement $P \leftrightarrow R$, as expected by duality.  The effect of the coupling between colors is to simply to shift the convolution in the flow equation.  
If the initial distributions are equal, the fixed point coupling distributions are of the same form as in the Ising case, and the rescaling of energies by $\Gamma$ makes the $\ln{\kappa}$ term irrelevant once one is below an energy scale on the order of $\Omega_I/\kappa$~\cite{Senthil:1996,Carlon:2001}.  This is clearly true if $\epsilon < \epsilon_c(N)$ initially. Then $\epsilon$ is lowered towards zero, hence $\kappa$ is pushed down towards 1 and $\ln \kappa \to 0$. We see, therefore, that for any finite value of $N$, the criticality of the  system is that of the infinite randomness Ising fixed point.
Note that from symmetry $\kappa$ does not renormalize for either the clock model or the Potts model~\cite{Senthil:1996}, unlike the present case. So, the above analysis is all that we need to perform.

If the initial value of $\epsilon$ is larger than $\epsilon_c(N)$,  $\kappa$ grows without bound, so the energy cutoff below which one must be to observe universal scaling behavior of RTFIM is driven to zero; in other words, the strong disorder renormalization group analysis breaks down. This breakdown of scaling may imply persistence of $\mathrm{QPT_{1}}$ and will in turn imply an important modification of the Aizenman-Wehr theorem. The conjecture is currently being checked in numerical simulations~\cite{Jia:2007}.

Consider a Hamiltonian, $H=H_{0}+g H_{1}$, where $H_{0}$ and $H_{1}$ commute.  A level crossing can take place at $g_{c}$, where an excited state drops below the ground state at $g_{c}$. This will correspond to a first order transition and is possible even in a finite system. An example is a metamagnetic transition tuned by an external magnetic field~\cite{Aeppli:2001}. Our work cannot be relevant to this problem, as the thermodynamic limit was essential for the argument given above regarding the rounding of $\mathrm{QPT_{1}}$ by disorder into a continuous phase transition. It is easy to see that in this case the disorder will merely  broaden the transition. By contrast, the problem we considered involved non-commuting $H_{0}$ and $H_{1}$, and the $\mathrm{QPT_{1}}$ was driven by quantum fluctuations.

A striking but simple extension that may also find applications to numerous  complex strongly correlated systems such as organics, heavy fermions, and high-$T_{c}$ superconductors is when the $\mathrm{QPT_{1}}$ in the pure problem  is between two ordered states, which from Landau theory is generically a first order transition. The heuristic argument goes through straightforwardly if we are mindful that both sides of the transition involves broken symmetries, albeit of different types.

We thank J. Lebowitz, G. Murthy, J. Rudnick, R. Shankar, T. Vojta, and A. P. Young  for important comments.This work was supported by the National Science Foundation, Grant. No. DMR-0705092. S. C. would also like to thank the Aspen Center for Physics.

\end{document}